# Theoretical study on charge transfer properties of triphenylamino-ethynyl Polycyclic Aromatic Hydrocarbon derivatives

**Zhipeng Tong [a†] ,Xiaoqi Sun [a,c†],Guiya Qin [b,d] , Jinpu Bai [a] ,Qi Zhao [a] ,Aimin Ren [b] , and Jingfu Guo [a] ***

[a] College of physics, Northeast Normal University, Changchun 130024;

[b] Institute of theoretical chemistry, College of Chemistry, Jilin University, Changchun 130023

[c] Department of Physics, College of Science, Yanbian University, Yanji 133002

[d]Chemistry Teaching and Research Section, Department of Basic Sciences, Jilin Jianzhu University, Changchun, 130118

*Corresponding author：Guo Jing-fu, E-mail address: guojf217@nenu.edu.cn.

†equally contributed to this work

**Abstract** The regulation mechanisms of backbone topology (tri-/tetracyclic arenes), substitution positions, and functional groups on charge transport properties are systematically investigated for triphenylamine-ethynylene fused acene derivatives by DFT combined kinetic Monte Carlo simulations and Marcus charge transfer theory in this study. The results demonstrate that sulfur-doped tricyclic arene backbones (benzodithiophene and anthracene) effectively suppress high-frequency vibrational modes (1191 $cm^{-1}$), reducing reorganization energy to 146.1 meV. Concurrently intermolecular π-π slippage (3.7-4.0Åspacing) also is optimized, hence enhancing 2D hole mobility ($\mu_{2D}$=0.18$cm^2V^{-1}s^{-1}$). Notably, asymmetric charge transport pathways in 2,7-disubstituted pyrene(27DTEP) decrease transfer integrals by 34%, while 1,6-substitution (16DTEP)reconstructs HOMO distribution and induces rotational stacking (θ =79°), boosting transfer integrals by 28% and improving isotropy of mobility. Further a "backbone-functional group synergy" is proposed, revealing that concentrated HOMO distribution localized on the backbone (≥62.4%) amplifies transfer integral gains(ΔV= 9.1 meV), outweighing the 38% increase in reorganization energy and significantly enhancing mobility. These findings provides theoretical basis and quantitative model for the rational design of high-mobility organic ultraviolet photodetectors.

# 基于三苯胺乙炔基稠环芳烃衍生物电荷传输性质的理论研究

佟志鹏[1†]，孙晓琦[1, 3†]，秦桂亚[2, 4]，白晋朴[1]，赵琦[1]，任爱民[2]，郭景富[1*]

[1] 东北师范大学物理学院，长春 130024

[2] 吉林大学化学学院理论化学研究所，长春 130023

[3] 延边大学，理学院物理系，延吉 133002

4 吉林建筑学院基础科学学部化学教研室，长春 130118

*Corresponding author：Guo Jing-fu, E-mail address: guojf217@nenu.edu.cn.

†equally contributed to this work

摘要：本研究基于三苯胺乙炔基并苯衍生物的分子设计，通过 Marcus 电荷传输理论结合动力学蒙特卡罗模拟，系统探究了骨架拓扑结构（三环/四环芳烃）、取代位点及取代基类型对电荷传输性能的调控机制。研究结果表明，三环芳烃骨架（苯并二噻吩、蒽）通过硫原子掺杂可有效抑制高频振动模式（1191cm$^{-1}$），将重组能降低至 146.1 meV，同时优化分子间π-π滑移（间距 3.7-4.0Å），实现空穴迁移率提高（$\mu_{2D}$=0.18cm$^2$V$^{-1}$s$^{-1}$）。四环芳烃芘骨架（27DTEP）在 2,7 位取代时因传输通道不对称导致转移积分降低 34%，而 1,6 位取代重构 HOMO 轨道且使二体分子间堆叠产生旋转（θ=79°），使（16DTEP）转移积分提升 28%并显著增强迁移率的各向同性。进一步提出“骨架-取代基协同调控”策略，阐明当轨道分布集中于骨架时（占比≥62.4%），虽重组能增大 38%，但转移积分增益（ΔV=89.1 meV）更大，致迁移率显著提升。本研究为高迁移率有机紫外光电探测材料的理性设计提供了理论依据与量化模型。

# 1 引言：

有机半导体在科技发展中扮演着举足轻重的角色，从小到几纳米的芯片手机到大到航天器等高科技产品都与有机半导体的发展有着密不可分的关系。有机半导体材料具有良好的延展性[1]，可大面积加工[2,3]，质量较轻且价格低廉[4,5]等特点被广泛应用于有机场效应晶体管（OFETs）[6,7]、有机发光二极管（OLED）[8,9]、有机光伏器件（OPV）[10,11]等电子器件中。近年来，随着紫外探测技术在导弹预警系统、保密光通信及皮肤癌筛查等军事和民生领域展现出不可替代的应用价值，基于有机半导体的紫外光电探测研究已成为先进光电材料领域的前沿课题。其在装备轻量化、柔性可穿戴器件集成等方面的独特优势，正推动着新一代探测技术向高灵敏度、低功耗和环境适应性方向突破。[12]。一般来说，有机光晶体管（organic phototransistors(OPTs)）工作原理如下：有机半导体吸收光子并产生激子；随后激子扩散并解离成自由载流子；自由电荷载流子流向相应的电极并导致电流信号的变化[13]。所以，良好的紫外有机光电探测器材料不仅要有紫外区域的独特吸收，还要有较高的迁移率和相应的发光量子效率[14]。天津大学胡教授课题组近年来通过给体-受体协同调控策略，成功设计并合成了 BBDTY[15]与 oF-PTTTP[16]两种π共轭有机半导体材料。实验表征显示，二者不仅展现出优异的紫外光响应特性（响应度达 857A/W），其空穴场效应迁移率（$\mu_h$=0.12cm$^2$V$^{-1}$s$^{-1}$）较传统并苯类材料提升 2 个数量级。值得注意的是，该迁移率仍显著低于理论极限值(>10cm$^2$V$^{-1}$s$^{-1}$)，揭示出分子的迁移率还有较大的提升空间。吉林大学田教授课题组合成的系列三苯胺基给体-受体-给体（D-A-D）型结构的分子 48DTEBDT[17],27DTEP, 16DTEP[18],TBA-An,TPA-An[19],不仅有很好的紫外光响应度（2.86×10$^6$→1.04×10$^5$AW$^{-1}$）、探测度（1.49×10$^{18}$→5.28×10$^{16}$ Jones),给体(D)三苯胺基和不同受体(A)交替组成带来了较高的发光量子产率，但载流子迁移率相差较大（$\mu_h$=0.025-2.1 cm$^2$V$^{-1}$ s$^{-1}$,），且在载流子迁移率上也存在较大的提升空间。值得关注的是，取代基三苯胺基由于其独特的螺旋桨结构、富电子性质、低电离电位、宽吸收带在有机染料太阳能电池、有机场效应晶体管等领域备受青睐[20-22]。然而上述系列宽带隙材料普遍面临π共轭有限导致的电荷传输效率低下问题，使得这类宽带有机半导体材料数量有限，这严重制约了其在 OPTs 器件中的应用。究其本质，材料性能瓶颈源于：（1）分子振动模式与重组能的非最优匹

配；（2）分子间转移积分受堆积模式制约。

本论文拟通过选出的系列高紫外光响应度分子（图 1）出发，重点探究该系列分子迁移率小的关键结构影响因素，总结不同骨架、不同位点及取代基改变对系列分子电荷传输性质的影响规律。保留三苯胺乙炔基取代基的高发光量子产率优势下，从振动模式、分子间轨道重叠的角度来探究 D-A-D 型系列分子的供体和受体分别对重组能、转移积分的影响；探究分子骨架、取代基不同改变如何影响堆积模式以及最大传输通道的转移积分，进而影响迁移率大小的事实，旨在为高性能有机紫外光电探测器材料的研发提供参考与见解。本研究的创新性在于：从分子振动-电子耦合的量子力学本质出发，建立多维度的结构-性能关联模型，为突破有机半导体"高响应-低迁移"困境提供理论指导。

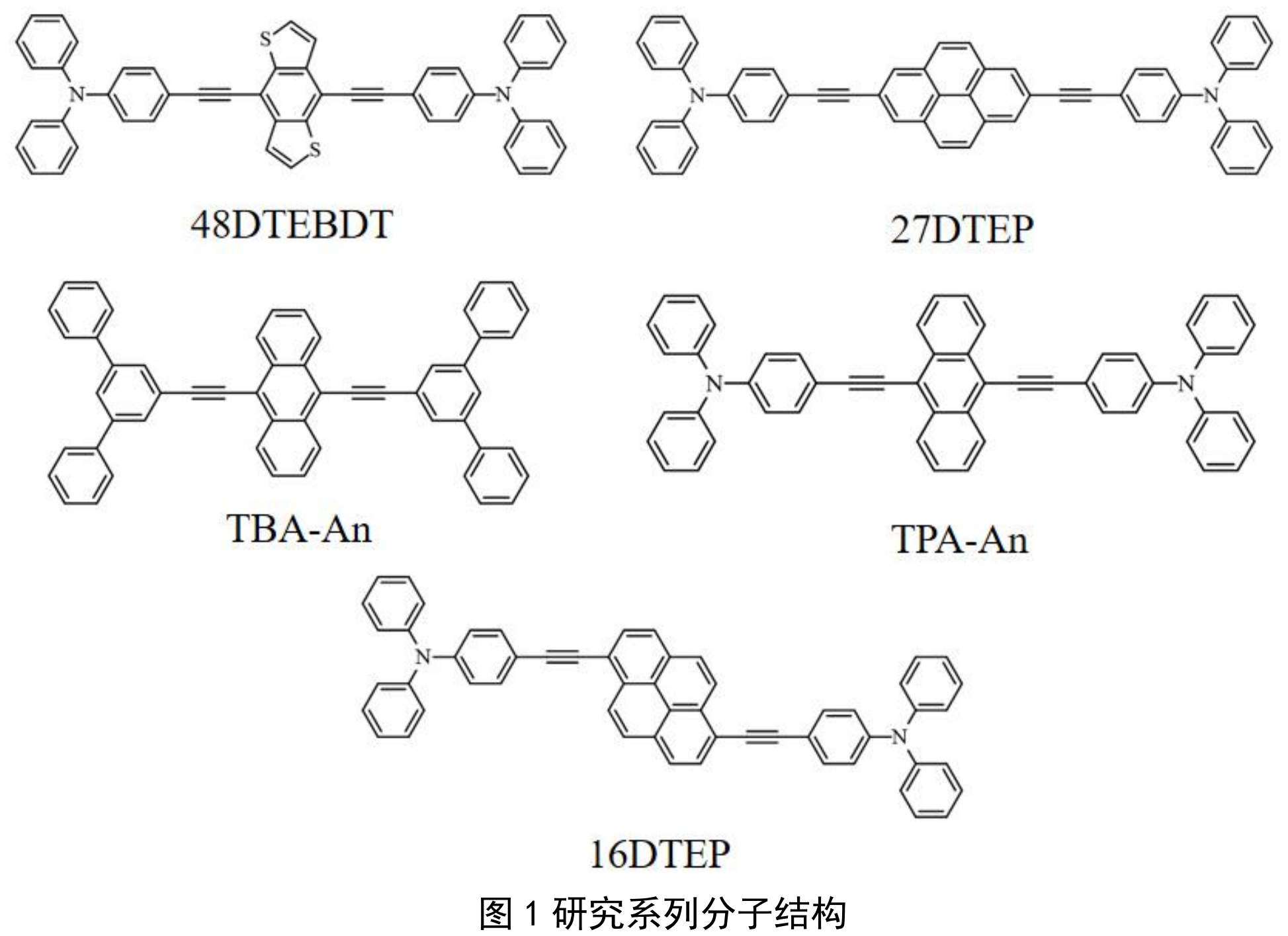

**图 1 研究系列分子结构**

## 2 计算方法：

有机半导体的电荷传输方式分为三种模型：能带模型[23]，Holstein-Peierls 模型[24]，跳跃模型[25]。由于转移积分远小于重组能，本研究体系的电荷传输归类于跳跃模型，因此可用 Marcus 公式描述。本文基于 Marcus 公式在 B3LYP/6-31G（d,p）水平下对系列分子的中性态和离子态的几何构型进行优化，在 Gaussian09 程序包中完成；基于 Marcus 理论的电荷转移公式[26,27]如下：

$$k_{CT}=\frac{1}{\hbar}|V|^2\sqrt{\frac{\pi}{\lambda k_BT}}\exp\left(-\frac{\Delta G_0+\lambda}{4k_BT}\right)$$

式中：$\hbar$为普朗克常数；$k_B$为玻尔兹曼常数；T(k)为温度；$\lambda$(meV)为重组能；$V$(meV)为电荷转移积分，对于电荷自交换情况$\Delta G_0$为零。固体中的电荷传输是一种布朗运动，因此本工作通过随机行走法来模拟晶体材料中的电荷扩散系数 $D$。利用课题组自编程序 CTMP 软件包[28]实现 2000 次随机行走模拟，获取均方位移$\langle x^2(t)\rangle$值，将其对扩散时间 t 作图拟合得到电荷扩散系数 $D$[29]，公式如下：

$$D=\frac{1}{2n}\lim_{t\to\infty}\frac{\langle x^2(t)\rangle}{t}$$

式中：n 为传输维度，x(t)是电荷随时间 t 扩散的距离利用爱因斯坦公式[30]计算载流子迁移率：

$$\mu=\frac{eD}{k_BT}$$

$e$ 为电子电荷;$k_B$为玻尔兹曼常数。

由于晶体中分子堆积模式不同，电荷沿着不同方向传输速率不同，因此存在迁移率的各向异性，理论上各向异性迁移率的计算公式如下[31]：

$$\mu_{2D}=\frac{e}{2k_BT}\lim_{t\to\infty}\sum\frac{x(t)^2\cos^2\gamma\ \cos^2(\theta_i-\Phi)}{t}$$

式中：$\gamma_i$ 是层间传输方向与参考轴的夹角，当仅考虑二维平面传输时，$\gamma_i$为 0；$\theta_i$ 是第 $i$ 个跳跃方向与参考轴夹角；$\Phi$为目标方向相对于参考轴的方位角。

# 3 结果与讨论：

## 3.1 分子几何结构、前线分子轨道、电离能

为了选择合适的理论计算方法，选择模型分子为有实验晶体结构的分子，初始分子结构取自剑桥晶体数据库。分别在气相和固相优化，将两种理论预测的模型结构与实际晶体结构对比，通过计算发现，气相下优化后分子的几何结构与实验上合成的晶体结构差异较大，基于固相下优化后结构的键角($C_2$-$N_1$-$C_1$)角度

-41° 而气相下优化后的为-46°，通过与实际晶体比较发现固相下优化的单分子结构与实验合成的晶体结构键角（$C_2$-$N_1$-$C_1$)为-35°）更为相似，因此全文讨论基于固相下优化结构进行。计算中使用量子力学/分子力学方法[32,33]（quantum mechanical/molecular mechanic，QM/MM）模拟固相环境下中心分子的中性态和离子态的几何构型，中心分子层使用 B3LYP/6-31G（d,p）水平，外层分子冻结并用 UFF 力场计算。

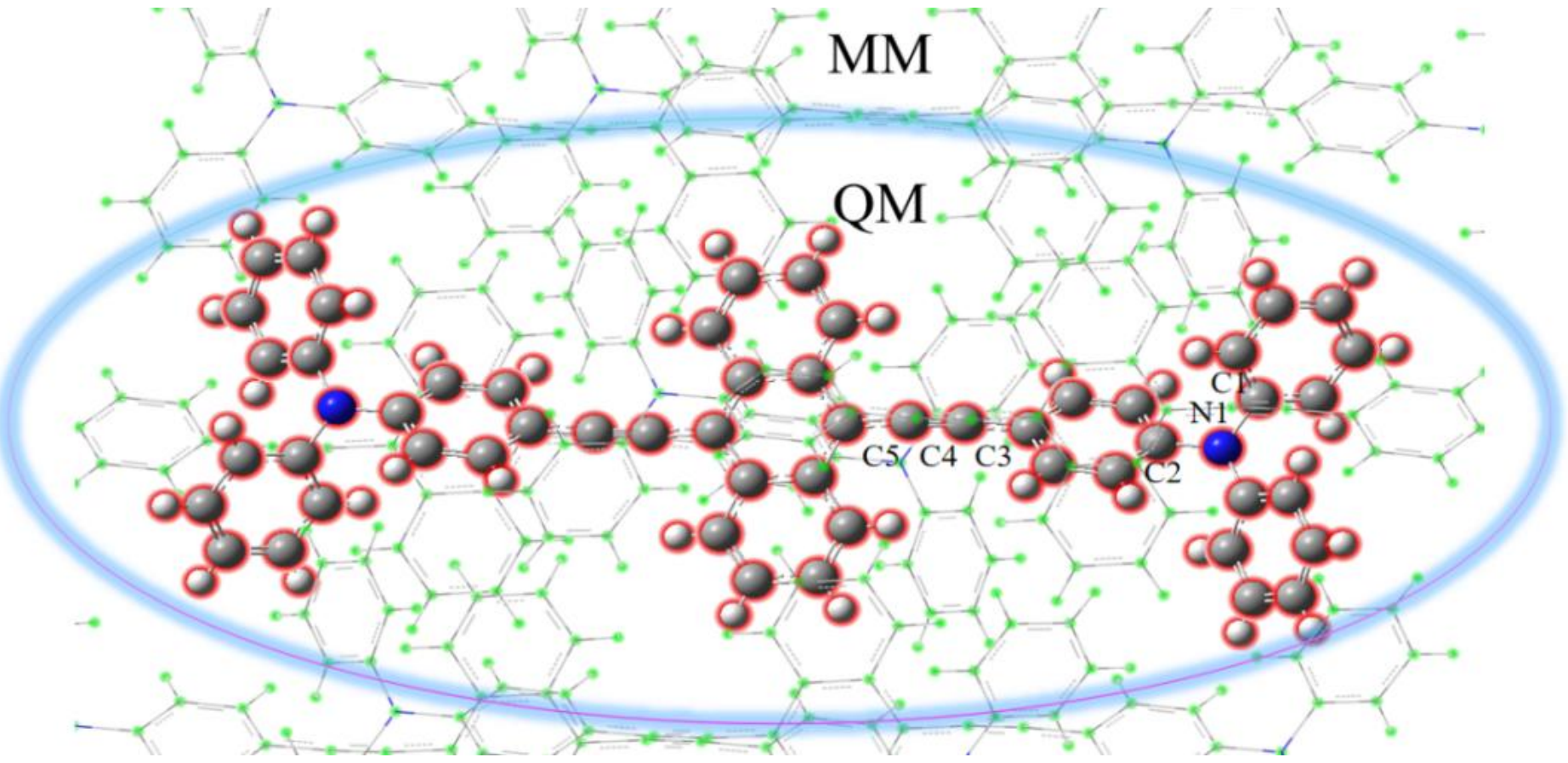

图 2 球棍模型（红色标识）表示的中心分子使用 QM 方法优化，用线性模式（蓝色标识）表示的周围相邻分子使用 MM 方法处理

器件的传输性能主要由两大因素决定，一是空穴的注入能力，另一个是载流子迁移率的大小。我们首先计算并探讨这系列分子的前线分子轨道来评估空穴的注入能力。在 B3LYP/6-311+G（d,p）水平下计算了系列分子的轨道能级，示于图 3。实验中常取金（功函为 5.1eV）用作电极。计算结果表明这系列分子 LUMO 值都较高，不利于电子的注入，这与实验仅测到空穴迁移率的结果相符合。由于这系列分子都是空穴传输, 因此主要分析了分子结构对 HOMO 能级调节情况。骨架为三环芳烃苯并二噻吩和蒽的两个分子 48DTEBDT 和 TPA-An 的 HOMO 能级分别为-5.06eV，-5.03eV，其绝对值相比于骨架为四环芳烃的分子 16DTEP、27DTEP 分子更接近电极功函 5.1eV，表明三苯胺基取代骨架三环芳烃比四环芳烃的空穴注入能力较强。相比 TPA-An，具有间三联苯取代基取代的 TBA-An 分子中，间三联苯乙炔基对 LUMO 能级影响较小，但对 HOMO 能级的降低比三苯胺乙炔基幅度大。对比 16DTEP、27DTEP 两分子，1,6 位点使得 HOMO 能级升高，LUMO 能级降低，这使得分子 16DTEP 的能隙较小，光稳定性大大降低。

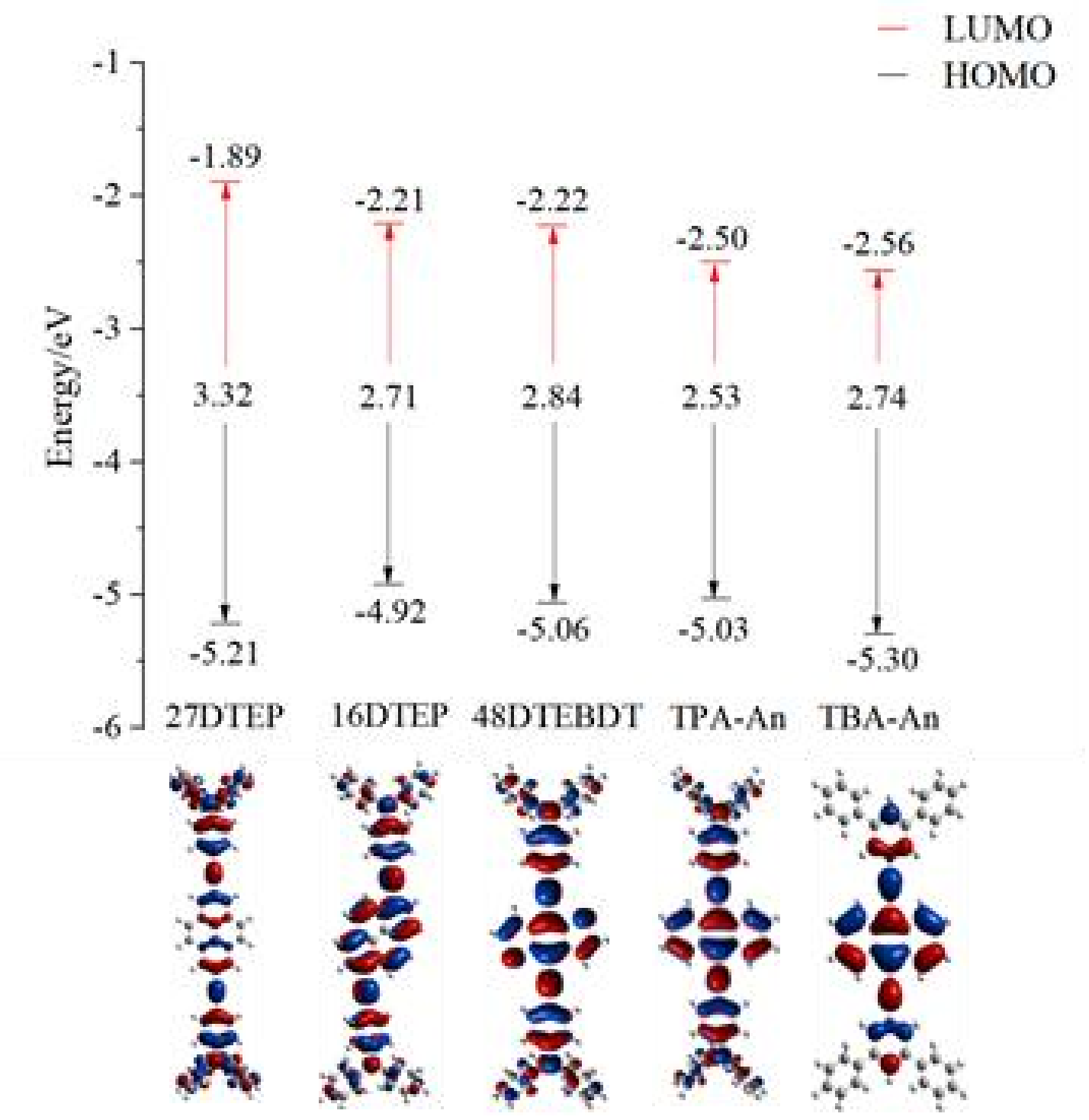

图 3 系列分子的 HOMO 和 LUMO 能级及 HOMO 轨道分布（单位：eV）

### 3.2 重组能、正则模式分析、键长变化分析

影响器件性能的另一个重要因素是载流子迁移率，有机半导体的迁移率主要与两个因素有关，一个是重组能，另一个是转移积分。首先我们通过绝热势能面（AP）法和正则模式方法（NM）分别计算了系列分子的空穴重组能，结果列于表 1 中。两种方法计算的结果相比较发现，AP 计算方法和 NM 计算方法得到重组能的大小基本相同，所以说该系列分子的热振动属于谐振子行为，符合谐振子近似。为分析方便，定义振动区域 0-1000 $cm^{-1}$ 为低频区域，大于 1000 $cm^{-1}$ 为高频区域，通过观察数据发现这系列分子的三个分子的骨架（苯并二噻吩、蒽、芘）的重组能振动主要分布在高频区域，重组能大小分别为 111.6 meV、140.3 meV、139.1 meV。

由表 2 可以发现三苯胺基取代的骨架三环芳烃分子 48DTEBDT、TPA-An 的重组能分别为 146.1 meV、182.5 meV，相比于骨架四环芳烃分子 27DTEP 的重组能（87.6 meV）具有较大的重组能。对比分子 16DTEP 和 27DTEP，不同位点 1,6 位点的取代使得重组能由 87.6 meV 变为 166.6 meV，重组能大幅度增加。而对比分子 TPA-An（182.5 meV）和 TBA-An（148.6 meV），间三联苯取代三苯

胺基使得重组能略有减小。为了更详细的探究其原因，对这些分子的正则振动模式进行了分析。

表 1 用 QM/MM 模型计算了这系列分子内空穴重组能（$\lambda_{AP}$/ $\lambda_{NM}$）和高、低频区（$\lambda_{low}$/ $\lambda_{high}$）的分布及垂直电离势/垂直亲和势（VIP/VEA）值。（单位：eV）

| Molecule | $\lambda_{AP}$/meV | $\lambda_{NM}$/meV | $\lambda_{low}$/meV | $\lambda_{high}$/meV | VIP | VEA |
|---|---|---|---|---|---|---|
| 48DTEBDT | 146.1 | 148.9 | 76.2 | 72.6 | 5.60 | 0.86 |
| 16DTEP | 166.6 | 173.5 | 50.8 | 122.6 | 5.46 | 0.82 |
| TPA-An | 182.5 | 180.2 | 94.5 | 85.6 | 5.56 | 1.06 |
| 27DTEP | 87.6 | 87.7 | 53.3 | 34.4 | 5.68 | 0.51 |
| TBA-An | 148.6 | 149.0 | 32.4 | 116.3 | 6.02 | 1.26 |

表 2 用 QM/MM 模型计算骨架内空穴重组能（$\lambda_{AP}$/ $\lambda_{NM}$）和高、低频区（$\lambda_{low}$/ $\lambda_{high}$）（单位：eV）

| Molecular backbone | $\lambda_{AP}$/meV | $\lambda_{NM}$/meV | $\lambda_{low}$/meV | $\lambda_{high}$/meV |
|---|---|---|---|---|
| **Benzodithiophene** | 165.2 | 167.1 | 55.5 | 111.6 |
| **Anthracene** | 139.1 | 141.5 | 0.5 | 140.3 |
| **pyrene** | 154.0 | 154.0 | 14.1 | 139.1 |

分析三苯胺乙炔基 2,7 位点取代的系列分子振动模式可以发现（见图 5），对于三环芳烃分子，当骨架由苯并二噻吩（48DTEBDT）变为蒽（TPA-An）时，低频 150.4cm$^{-1}$ 处的振动增大为 14.1meV，增大的振动为三苯胺基的摇摆振动；高频处 1596cm$^{-1}$ 的振动相比于骨架苯并二噻吩并没有较大的增强，但是却新增高频处 1199cm$^{-1}$ 较大的振动（8meV），振动是由蒽骨架摇摆振动和紧邻炔键取代基上的苯环的摇摆振动产生的，以上高低频区振动增强是 TPA-An 重组能较大的原因。骨架由三环芳烃蒽转换成四环芳烃芘（27DTEP）之后，骨架芘不仅降低了低频区的取代基振动，同时降低了高频区骨架自身产生的振动（见表 1）。与 27DTEP 相比较，1,6 位点取代虽然使得分子 16DTEP 在 156cm$^{-1}$ 处低频振动高达 9.7meV，但二者低频区总的振动重组能相差不大。而高频区的振动模式发生改变，27DTEP 中为骨架芘的摇摆振动，而 16DTEP 中为取代基参与的骨架拉伸振动。这使得 16DTEP 高频区的振动相比于 27DTEP 有所增大。而三苯胺乙炔基和间三联苯乙炔基取代蒽骨架（TPA-An，TBA-An）相比较发现，间三联苯基能有效降低低频的振动，这可能与间三联苯取代基的结构刚性有关（三苯胺基则

易发生空间旋转柔性大）。不同于三苯胺乙炔基的是，虽然间三联苯也使得高频区（546、1329、1608 $cm^{-1}$ 骨架伸缩振动）的振动有所增加，但它同时使低频区振动（500cm-1 以下振动几乎被抑制）更大程度降低，这是为什么间三联苯取代基使得 TBA-An 分子重组能比 TPA-An 明显减小的原因。为了更好的分析取代基和骨架与重组能的关系，也计算了骨架片段的重组能结果列于表 2 中。取代基为三苯胺乙炔基时，骨架为三环芳烃蒽（TPA-An）重组能增大，骨架为四环芳烃芘（27DTEP）重组能减小。通过分析 HOMO 组份发现当三苯胺基取代的骨架分别为三环芳烃和四环芳烃时，HOMO 组分发生了截然不同的变化，我们通过观察 2,7 位点取代骨架三环芳烃分子 TBA-An、TPA-An、48DTEBDT 的 HOMO 组份分布图（见图 7 和表 3），发现骨架的轨道分布占比依次逐渐减小，而且骨架的轨道分布占比在单分子中最大。

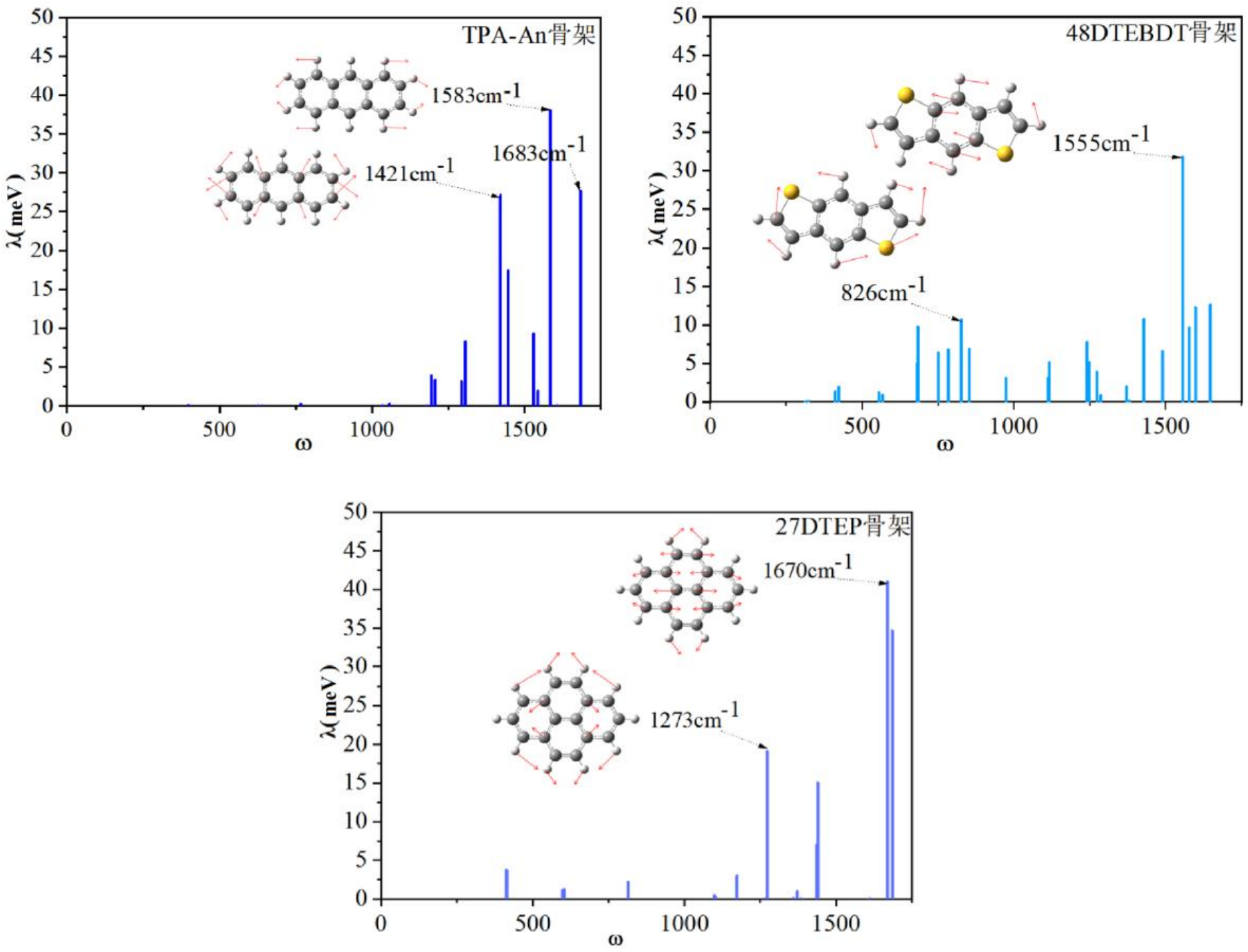

**图 4 研究分子中骨架振动模式对空穴重组能的贡献**

而 2,7 位点取代骨架为四环芳烃的芘分子 27DTEP 分子的轨道分布图的趋势则大为不同。骨架芘由于特殊的结构使其自身轨道分布占比小，而两端取代基上的轨道分布占比大，因此两端取代基的轨道分布趋势对重组能有着重要的影响。同时，具有相同的骨架和不同取代基的分子（如 TBA-An 与 TPA-An）相比较发现，TBA-An 中的间三联苯取代基使得骨架蒽在 HOMO 中的占比大，取代基自身占比较小且刚性强(取代基中二苯基几何对称），这也使得 TBA-An 分子的低

频振动比 TPA-An 分子更小，因此取代基本身的振动强度降低，但引起骨架在高频区的振动却有提高，这表明间三联苯导致骨架的振动强度增大，表现出与三苯胺基乙炔基对同一骨架的影响大不相同。

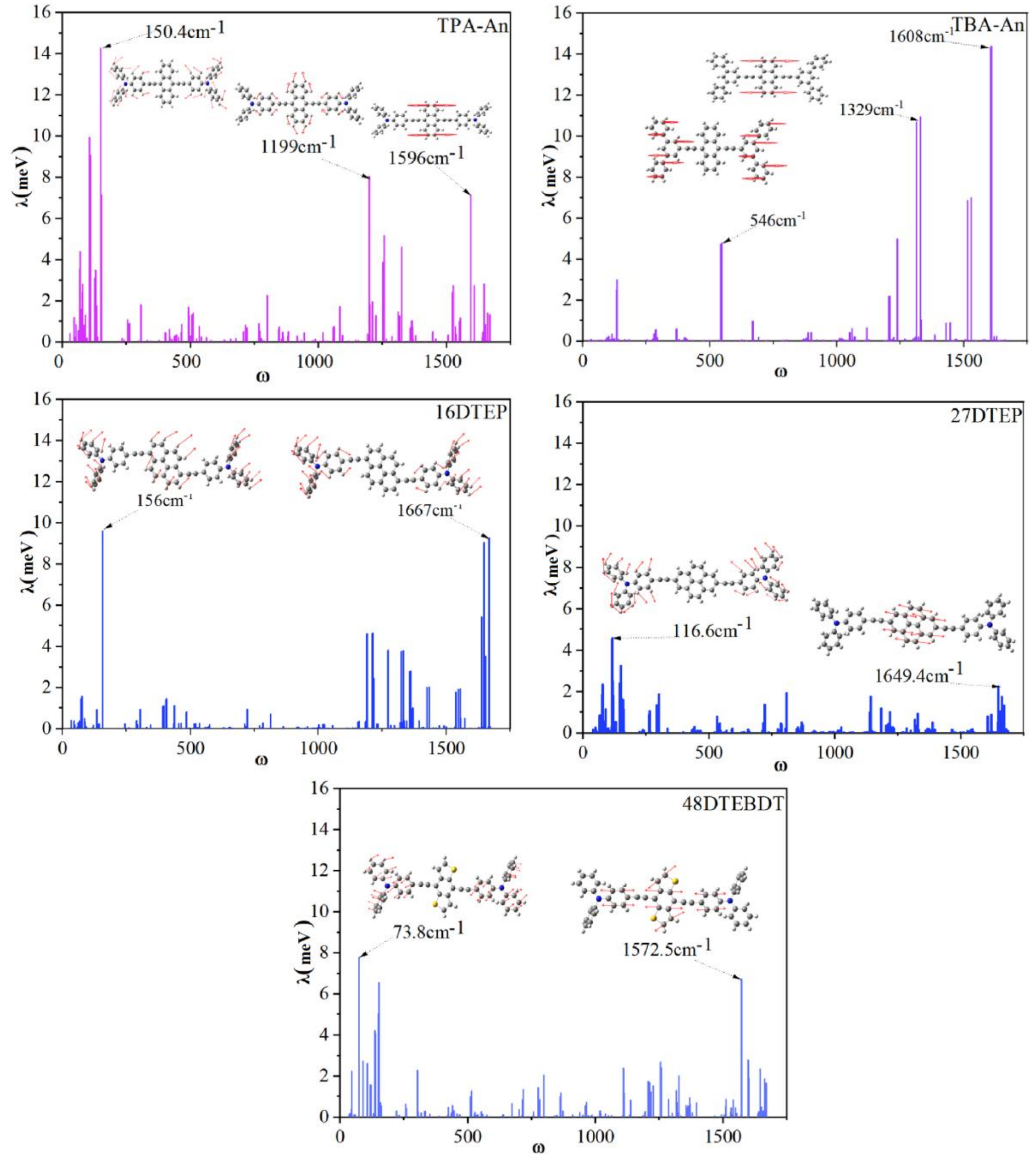

图 5 振动模式对所研究各分子空穴重组能的贡献

为了深入分析取代基与骨架对重组能的影响，我们对空穴传输时键长变化进行了分析。通过计算分子得失电子后中性态和离子态下几何弛豫的变化，能够很直观的反映出上述观点。下面统计分析了大于等于 1%的键长变化值对分子的重组能的贡献。

我们通过从图 6 中可以发现系列分子的 20，22，24 号键变化较大对重组能有贡献，三个键分别位于三苯胺基中 C-N 单键和连接炔键的两个单键上。三苯胺基取代三环芳烃的苯并二噻吩和蒽时（图 6a），TPA-An 相比于 48DTEBDT

靠近取代基的 20 号键和靠近骨架的 24 号键的变化都较大，这也是 TPA-An 的重组能较大的原因。而三苯胺基取代的三环芳烃蒽 TPA-An 和四环芳烃芘 27DTEP 分子相比较发现（见图 6c），四环芳烃 27DTEP 的 20 号，22 号，24 号键的键长变化均减小，四环芳烃骨架能更有效抑制自身的振动，但对取代基振动的抑制效果较低。间三联苯取代三苯胺基后，通过对比 TBA-An 与 TPA-An 的分子的键长变化发现（见图 6d），间三联苯取代基的 20 号键变化率大大降低尽管靠近骨架上 24 号单键变化则有所升高，这使 TBA-An 重组能明显大于 TPA-An。而 1,6 位点取代骨架四环芳烃芘时（见图 6b），16DTEP 分子中骨架片段在 HOMO 中占比较大，这样的 HOMO 分布与三苯胺基取代三环芳烃的分子相似，通过上述分析及图 3 的 HOMO 轨道分布图发现，键长的变化大小趋势与分子 HOMO 的轨道分布有关。27DTEP 分子中越靠近骨架的键长变化对重组能贡献就越小，而观察分子的轨道分布发现，越靠近骨架的键长上轨道分布越小。那么 HOMO 轨道中骨架片段和取代基片段的轨道占比与其对重组能有何关系？为此，我们对取代基和骨架在 HOMO 轨道中的占比进行了计算，并通过骨架和取代基轨道分布占比比值大小与重组能的相关性进行了分析。

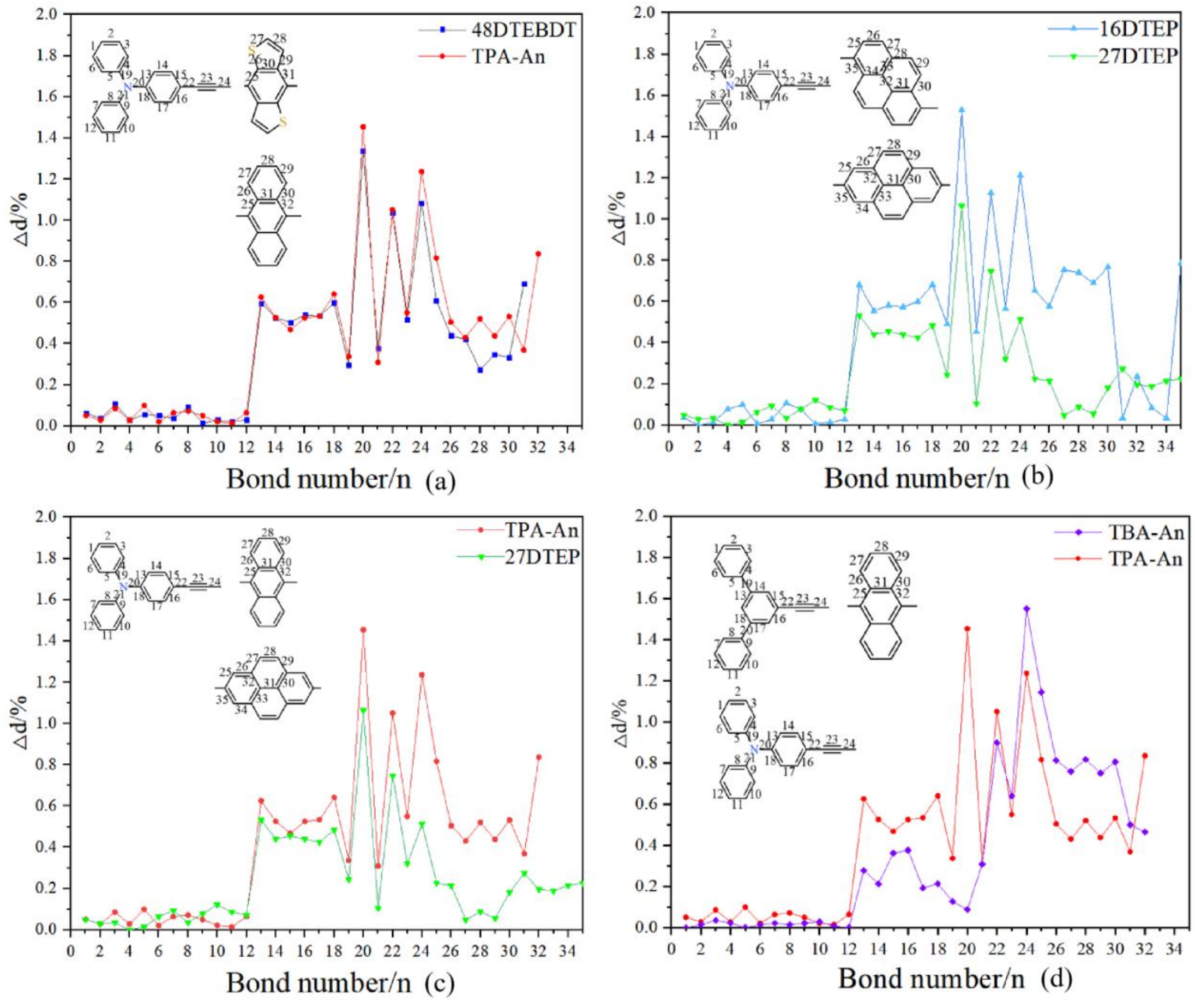

图 6 所研究各分子的键长变化图

### 3.3 轨道分布占比分析：

通过 Multiwfn 软件分析了这系列分子的骨架和取代基在 HOMO 中的分布比例，如图 7 和表 3 所示。HOMO 计算是在 B3LYP/6-311+G（d,p）水平下完成的。由表 3 可知，对于 27DTEP 分子，HOMO 轨道中的骨架片段组分与取代基片段组分的比值最小，其分子的重组能也是三苯胺乙炔基取代系列分子中最小，而且发现，骨架与取代基的轨道分布占比比值的大小趋势与其重组能大小趋势完全相同，这也验证了图 6 键长变化分析的结论，即重组能大小与骨架片段在 HOMO 中的轨道占比大小呈正相关，HOMO 中骨架组分越高，其空穴重组能越大。具有间三联苯乙炔基的 TBA-An 分子的骨架在 HOMO 中占比很大但重组能不符合上述趋势，原因是三苯胺乙炔基与间三联苯乙炔基对骨架影响的机制不同，间三联苯基对骨架的振动有较大的促进作用，同时抑制自身的振动。而三苯胺乙炔基则振动较强？

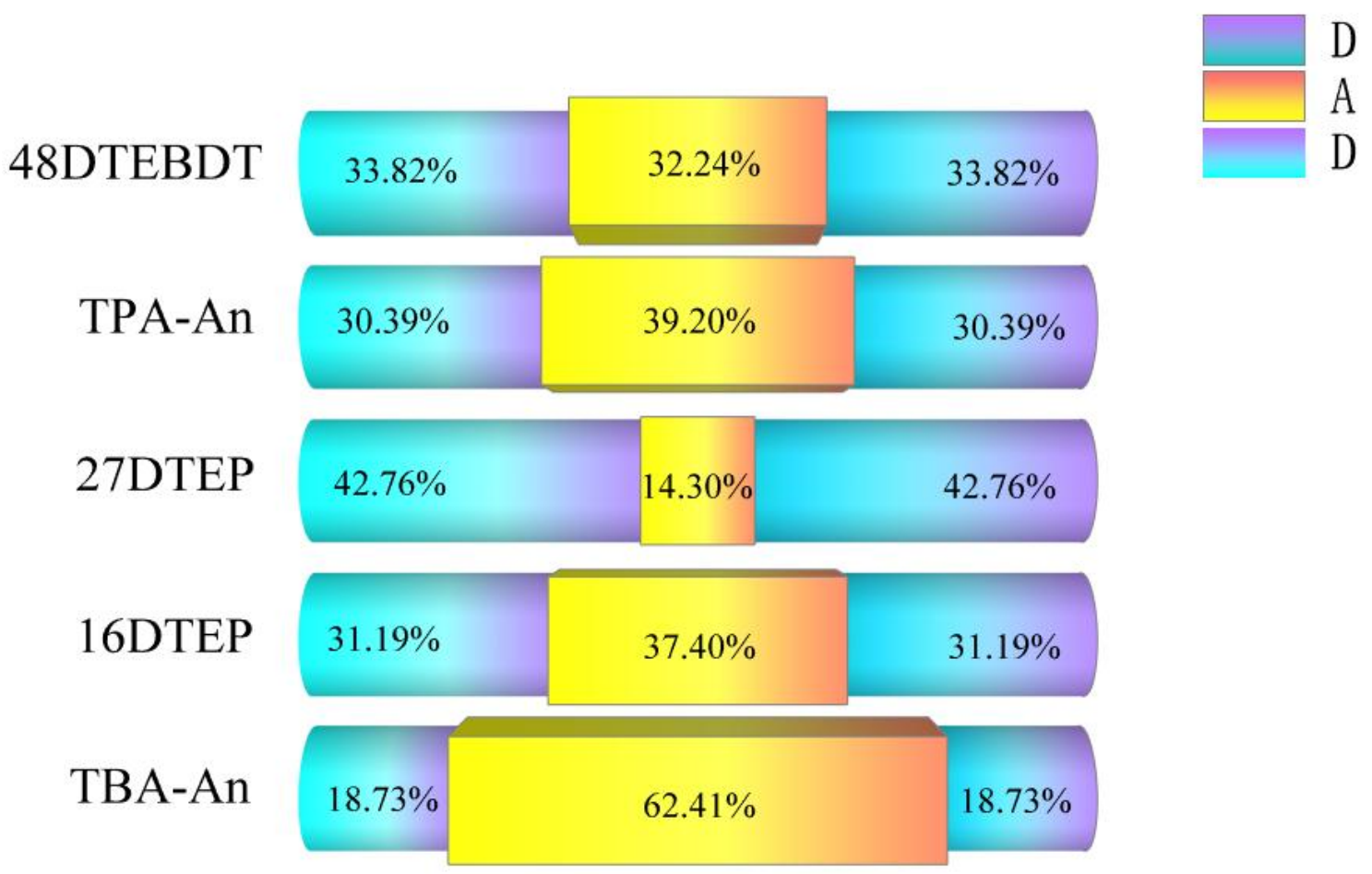

图 7 计算分子 HOMO 轨道中骨架片段和取代基片段的贡献比例

表 3 在 B3LYP/6-311++GDP 水平下计算的各研究分子骨架、取代基片段在 HOMO 中的占比和重组能关系

| 6-311++GDP | backbone | substituents | backbone/substituents | reorganization energy |
|---|---|---|---|---|
| 48DTEBDT | 32.24% | 33.82% | 0.95 | 146.1 |
| 16DTEP | 37.4% | 31.19% | 1.20 | 166.6 |
| TPA-An | 39.20% | 30.39% | 1.29 | 182.5 |
| 27DTEP | 14.30% | 42.76% | 0.33 | 87.6 |
| TBA-An | 62.41% | 18.73% | 3.33 | 148.6 |

### 3.4 分子堆积、分子间转移积分和弱相互作用：

转移积分是影响电荷转移速率的另一个重要因素，转移积分的大小和传输方向取决于分子的堆积排列模式和取向，系列分子的晶体堆积如图 8 所示。任选一个分子作为中心分子考虑由中心分子向周围最近邻分子的所有可能跳跃路径，计算了每个可能路径二体间的转移积分。对于研究分子的各个路径转移积分计算值及其质心距汇总在表 4 和图 8 中。分析图 8 中堆积模式可以发现：相比于骨架四环芳烃分子 27DTEP 的π-π砖状堆积，三苯胺基取代三环芳烃骨架 48DTEBDT 和 TPA-An 堆积的方式转变为人字形堆积，分子间最大传输通道的转移积分有所增大。而相比 TPA-An，间三联苯基取代后形成的 TBA-An 晶体中堆积方式不改变，但使得最大传输路径的转移积分有了显著的增加。这表明取代基的改变并不会对堆积方式有较大的影响，影响堆积方式的是骨架的变化。

对于三环芳烃骨架分子来说，骨架为苯并二噻吩与蒽的分子相比较，这两个分子的 HOMO 主要分布在骨架上，显然面对面的分子间的电子耦合最大（质心

距为 8.6Å和 7.6Å）是转移积分最大的传输通道；而四环芳烃骨架分子 27DTEP 的 HOMO 主要分布在取代基两端，27DTEP 晶体中的最大传输通道是质心距为 10.0Å的通道（见表 4 和图 8），该通道二聚体是边对边的分子堆积。与 27DTEP 相反，1,6 位点取代的 16DTEP 分子，其 HOMO 主要分布在骨架上，其晶体最大传输通道则是质心距离为 8.9Å的二体，是面对面的分子对。

众所周知，转移积分非常敏感于分子堆积的滑移和垂直距离的改变，因此我们归纳了最大传输路径中的二聚体的长短轴滑移和分子间距列于图 9 中。如图 9 所示，与 48DTEBDT 相比，骨架不同的 TPA-An 的长轴滑移有所增大（从 6.6 变为 7.7 Å），二聚体间的间距略有增大（从 3.7 变为 4.0 Å），使得 TPA-An 的传输积分小于 48DTEBDT。

表 4 所研究分子的空穴转移积分 (V/meV)，以及相邻二聚体之间的质心距离

| 48DTEBDT | | TPA-An | | 27DTEP | | 16DTEP | | TBA-An | |
|---|---|---|---|---|---|---|---|---|---|
| d | V（meV） | d | V | d | V | d | V | d | V |
| 7.6 | 29.2 | 8.6 | 10.4 | 4.9 | 1.6 | 8.9 | 8.4 | 4.8 | 96.8 |
| 7.6 | 29.2 | 8.6 | 10.4 | 5.0 | 0.4 | 8.9 | 8.4 | 5.0 | 89.1 |
| 9.0 | 0.7 | 8.7 | 1.8 | 9.8 | 2.4 | 8.9 | 8.4 | 10.5 | 2.9 |
| 9.0 | 0.7 | 8.7 | 1.8 | 10.0 | 5.2 | 8.9 | 8.4 | 10.5 | 2.9 |
| 9.4 | 7.1 | 9.5 | 4.5 | 10.0 | 5.2 | 9.3 | 0.4 | 11.2 | 2.6 |
| 9.4 | 7.1 | 9.5 | 4.5 | 10.3 | 0.2 | 9.3 | 0.4 | 11.5 | 0.8 |

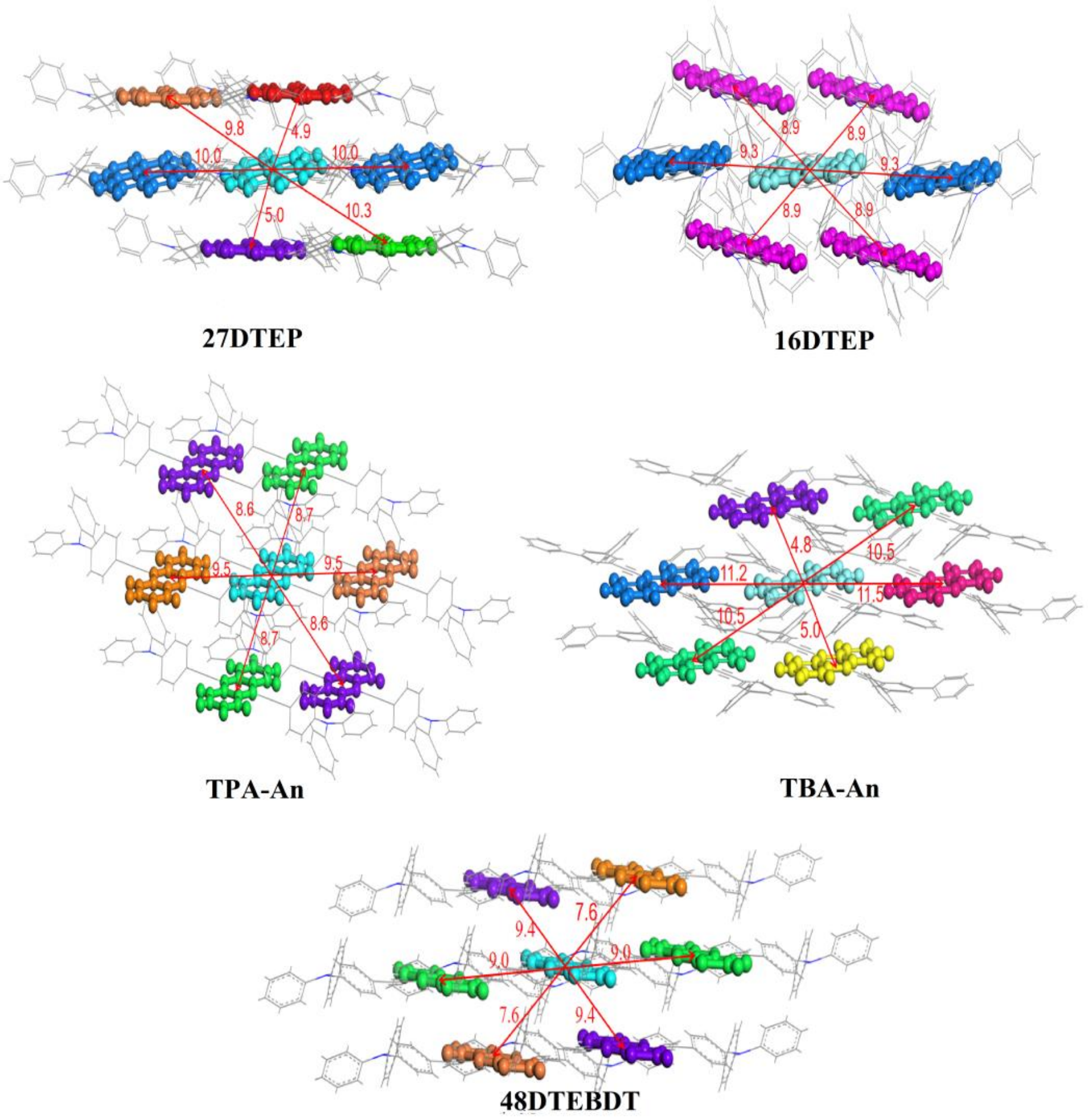

图 8 所有研究分子的晶体堆积模式和主要运输途径

而当骨架三环芳烃蒽转变为四环芳烃芘时，由于最大传输路径由面对面变为边对边传输方式，这使得 27DTEP 分子转移积分减小。之前的研究表明分子间采取面对面的砖型堆积是载流子传输较好的方式，与 2,7 位点相比较，1,6 位点取代骨架四环芳烃芘 16DTEP 的最大传输路径分子间为面对面堆积，转移积分有所增大，而 1,6 位点的改变同时带来分子间堆叠的相对旋转(79°)，这使得分子间的传输有了更多的对称通道，这是实验测得 16DTEP 的迁移率较大和二维各向同性较好的原因（在 3.6 和 3.7 节中进一步分析）。而间三联苯相比于三苯胺基缺少 sp3 杂化氮原子的连接使得取代基的空间位阻减小，这种变化不仅使分子间的滑移大幅度减小（7.7 Å→3.6 Å），分子间距减小（4.0 Å→3.3 Å），TBA-An 分子的最大传输通道的转移积分增大。为了进一步探究分子堆积变化对转移积分影响的根本原因，我们还计算了最大传输路径对 dimer 之间的 HOMO 重叠积分。

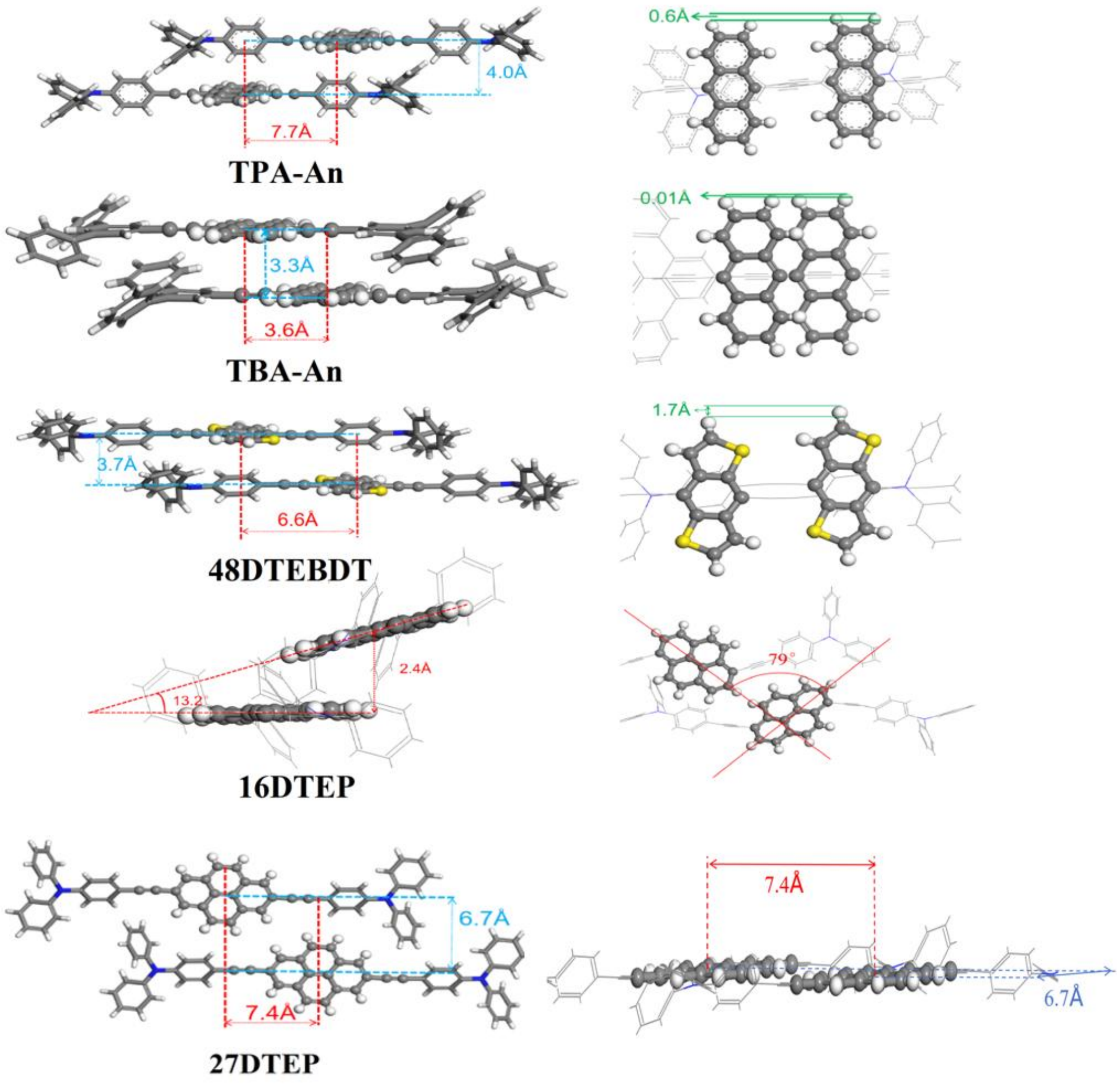

图 9 所研究的分子晶体面对面堆叠的二聚体滑移

分子间轨道重叠积分与前沿分子轨道相位分布有着密切不可分的关系，一般情况下两个单体 HOMO 间的重叠度与空穴转移积分呈正相关。二个单体 HOMO 间的重叠积分 Se 是由正相位积分（S+）和负相位积分（S-）的加和值得来的。我们发现三苯胺基乙炔基作为取代基时，不同的三环芳烃骨架分子相比较，相比于 TPA-An 的蒽，48DTEBDT 中的分子骨架苯并二噻吩会增大负向积分（负向积分 $S_-$ 由 0.00401 增大到 0.00669），同时使得正向积分减小（正向积分 $S_+$ 由 0.00529 减小到 0.00344），这使得 48DTEBDT 相比于 TPA-An 有较大的空间重叠积分，所以在最大传输通道有更大的转移积分。三苯胺基取代四环芳烃骨架分子 27DTEP 的空间重叠积分 Se=0.00053 比三环芳烃骨架 TPA-An（Se=0.00128）和 48DTEBDT（Se=0.00328）小一个数量级，所以这是取代骨架四环芳烃的空穴转移积分比取代骨架三环芳烃小的原因。而间三联苯取代基的改变使得负相积分

从 $S_-$=－0.00401（TPA-An）增加到 $S_-$=－0.0206（TBA-An），负相积分增大了一个数量级，但正相积分略有增加($S_+$=0.00529 增加为 0.0077)。这使得空间重叠积分分布有了大幅度提高（Se=－0.00128 增大到－0.0129），导致 TBA-An 相比于 TPA-An 最大传输路径的转移积分增大。1,6 位点取代使得分子间最大传输通道的分子堆积由边对边模式转换回面对面模式，这使得 16DTEP 的正负向积分分布的值比 27DTEP 的值都增大了一个数量级，所以 16DTEP 的空间重叠积 Se=－0.00143 比 27DTEP 的 Se=－0.00053 增大，这也是四环芳烃芘骨架 1,6 位点取代较 2,7 位点取代最大传输通道转移积分较大的原因。

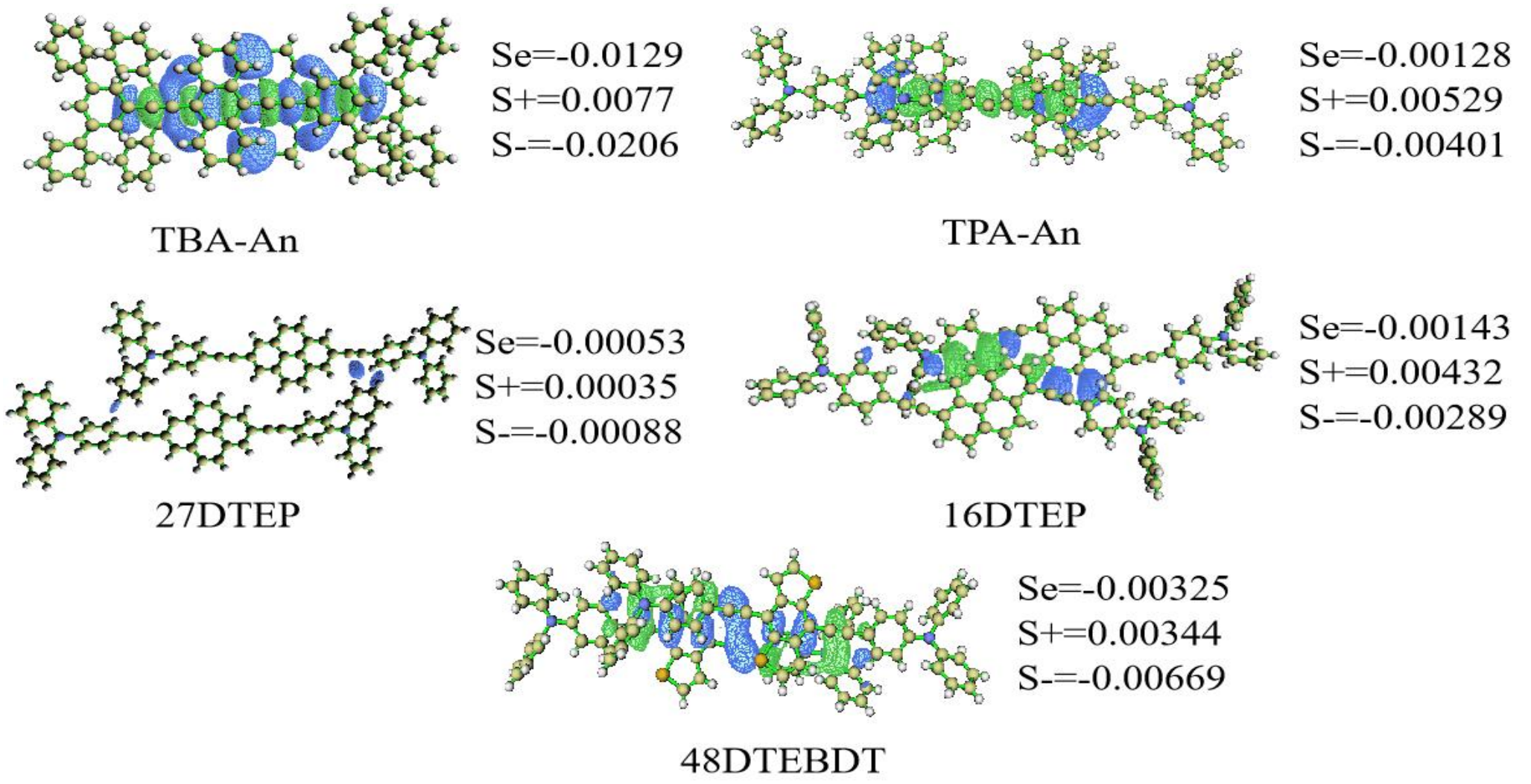

图 10 各分子间最大传输通道的轨道重叠积分

### 3.6 理论计算与实验计算迁移率：

载流子迁移率是衡量电荷传输性质的重要指标，模拟各分子晶体的迁移率结果列于表 7 中。通过对比发现，骨架三环芳烃取代的系列分子 48DTEBDT、TPA-An、27DTEP 的一维迁移率值比四环芳烃分子有所提高，分别为 1.01、0.16 和 0.12 $cm^2V^{-1}s^{-1}$。在实际测量中，一维堆积材料的电荷传输容易受缺陷及其测量条件的影响十分大，因此，测出的载流子迁移率往往和本征值相差较大。而具备各向同性载流子迁移率的分子材料由于传输通道不受限，制成器件后会有更好的性能体现。TBA-An 虽然有较大的一维迁移率，但根据上述分析表明，最大传输通道的转移积分虽然可观但并不对称，且分子具有较强的一维传输特性，较大的 IP 值，这些因素都使得空穴的注入能力变差，所以实验测得的迁移率小于理论计算值。16DTEP 虽然最大传输通道的转移积分相比 TBA-An 较小，但却有更多的对称传输通道，这使得传输方式由一维传输方式转变为二维传输方式，且 IP

值在这系列分子中最小，具有较强的空穴注入能力，因此实验测得的迁移率高。TPA-An 和 48DTEBDT 的情况与 16DTEP 一样。27DTEP 的本征迁移率较小，迁移率值基本与实验测量是一致的。

表 7　模拟了所有分子的空穴迁移率和实验值测得迁移率

| Molecule | $\mu_{1D}/(cm^2\cdot V^{-1}\cdot s^{-1})$ | $\mu_{2D}/(cm^2\cdot V^{-1}\cdot s^{-1})$ | $\mu_{exp\,(ave)}/(cm^2\cdot V^{-1}\cdot s^{-1})$ |
|---|---|---|---|
| 48DTEBDT | 1.010 | 0.180 | 0.250(0.160) |
| TBA-An | 2.000 | 0.800 | 0.150(0.070) |
| TPA-An | 0.160 | 0.024 | 0.450(0.240) |
| 16DTEP | 0.080 | 0.027 | 0.400(0.200) |
| 27DTEP | 0.120 | 0.023 | 0.025(0.012) |

3.7　各向异性迁移率：

为了更好地了解分子晶体中不同方向电荷传输性质，我们对研究分子晶体的二维各向异性迁移率进行了计算，结果显示在图 12 中。通过对各向异性迁移率的计算，三苯胺基取代四环芳烃分子比取代三环芳烃分子有更强的各向异性，骨架为三环芳烃中苯并二噻吩的取代极大的增强了分子的各向同性。间三联苯的取代使得分子的各向异性略有增强，1,6 位点的取代不仅能够使得传输方式由原来的一维传输转换为二维传输方式，也能使得分子各向同性略有增强，其各向异性迁移率具有最大值（0.10 $cm^2V^{-1}s^{-1}$）。这也是 16DTEP 分子器件有良好性能的原因。

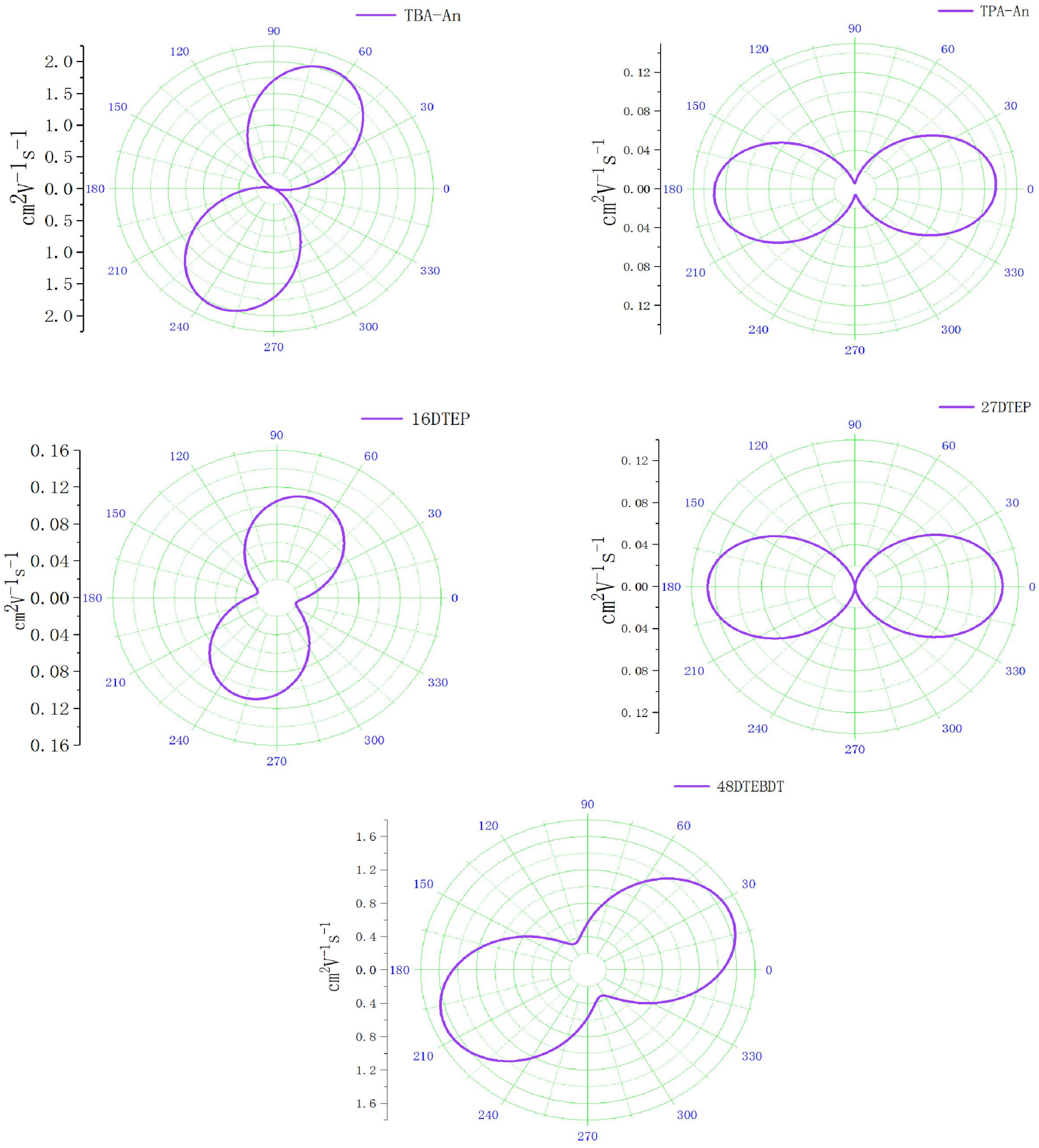

图 12 模拟了所研究分子的二维各向异性电子迁移率

## 4 结论

本研究选取三苯胺乙炔基供体-受体-供体型系列分子，采用 Marcus 电荷转移理论与动力学蒙特卡罗模拟相结合的方法，系统探究了并苯类衍生物的骨架结构（三环/四环芳烃）及取代位点不同及取代基不同对电荷传输性能的影响机制。研究发现，三环芳烃骨架（苯并二噻吩、蒽）通过硫原子取代可有效降低 1191 $cm^{-1}$ 高频振动重组能，同时改善分子间滑移距离，使 48DTEBDT 分子获得较大迁移率 0.18 $cm^2V^{-1}s^{-1}$，也使得分子的各向同性增强，所以用其制成的光电探测器的性能较高。四环芳烃芘骨架在 2,7 位取代时（27DTEP）虽能将重组能降低至 0.087eV，但 HOMO 分布向取代基偏移导致转移积分降低 34%。通过 1,6 位取代成功调控 HOMO 轨道分布于骨架上，实现转移积分提升 28%和二维各向同性迁移率的提高，使得 16DTEP 分子制成器件后载流子迁移率更优秀的表现，这与实验测试结果接近。

理论分析表明：（1）前沿分子轨道分布与重组能和转移积分大小具有一定

相关性。以三环芳烃为骨架的分子表现为二维砖状堆积，轨道重叠集中于骨架，形成有利于空穴传输的面对面堆积模式;四环芳烃为骨架的分子变为人字形堆积和 pitched-π堆积，通过取代位点调控可实现轨道分布从取代基向骨架的定向转移。（2）建立的单分子结构与电荷传输性能的关系模型表明，随着轨道分布在骨架占比的提高，重组能可能会提高，但转移积分也会提升。合理调节前沿轨道分布占比是调节转移积分和重组能的关键，也是增大载流子迁移率的途径之一。（3）对比 27DTEP 和 16DTEP 分子的传输各向异性表明，1,6 取代位点对各向同性的增大较明显,这有利于器件传输性能的提升。本工作提出的“骨架-取代基协同调控”策略为高迁移率有机半导体材料设计提供了理论依据。

**参考文献：**

$=$